\documentclass[a4paper,10pt]{article}
\usepackage[utf8]{inputenc}
\usepackage{graphicx}
\usepackage{epsfig}

\title{Energetic emissions from deconfinement in compact stars and their relation to the critical end point in the QCD phase diagram}
\author{D.~E.~Alvarez-Castillo\\
Bogoliubov Laboratory for Theoretical Physics,\\
Joint Institute for Nuclear Research,\\
Joliot-Curie Str. 6,
141980 Dubna,
Russia\\}

\begin{document}
\date{}

\maketitle

\begin{abstract}

In this work we study the case of deconfinement in compact star interiors in the presence of a strong first order phase transition associated to a critical end point in the QCD 
phase diagram. Neutron stars fulfilling these conditions show a third branch in the mass-radius diagram with the first and second branches being the white dwarfs and 
neutron stars configurations. The transition to the third branch can be reached by a pure hadronic neutron star through an 
induced collapse releasing energy that corresponds to a mass-energy difference between the second and third branch configurations. 
Physical outcomes of this phenomenon 
that can potentially explain the already detected astrophysical signals are discussed. In particular we present energy estimations for the case of a fast radio burst, 
seen as a double-peak structure in the object's light curve. 
\end{abstract}

\section{Introduction}

Neutron stars NS are evolved stars being created after the death of a massive star via a supernova explosion or a transition from a white dwarf accretion or dynamical instabilities.
Their interiors can reach up to several times the saturation density $n_0$, the canonical density inside atomic nuclei. It is quite uncertain what is the state of matter
under such high density conditions, therefore research on equation of state EoS is currently a very active area.
It is important to note that in NS temperature doesn't play a major role in the computation of the EoS, thus it can be neglected.
Recent observations have completely changed our understanding of the cold, dense nuclear matter in such compact star interiors. 

In this context, accurate mass determination has proved to be of great importance. 
In particular the observation of the 2M$_\odot$ pulsars, PSR J0348+0432~\cite{2013Sci...340..448A} and PSR J1614-2230~\cite{2010Natur.467.1081D},
have strongly constrained the stiffness of the NS EoS. On the contrary, radius measurements are not yet precise enough to test, discard, and select some of the many alternative EoS models. For instance,
frequency resolved pulse shape analysis for the nearest millisecond pulsar~\cite{2013ApJ...762...96B} supports relatively large radii 
while analysis from X-Ray bursters~\cite{Steiner:2010fz} point out to either moderate or short radii. Promising future radius measurements include upcoming
space missions, cf. NICER~\cite{nicer}.

Energetic phenomena like fast radio bursts FRB, - millisecond duration radio bursts from cosmological 
distances \cite{Lorimer2007} can be explained by compact stars by collapse into a black hole~\cite{2014A&A...562A.137F}. Under such scheme, 
the released energy of the process is $10^{40}$ erg. This scenario is not to be confused by early NS evolution consisting of hot protoneutron stars, with deconfinement transition in their 
interior that could serve as the mechanism triggering the core collapse supernovae~\cite{Sagert:2008ka}, as well as pulsar kicks at 
birth~\cite{2005ASPC..328..327P,2001ApJ...549.1111L,2005IJMPA..20.1148K,Berdermann:2006rk,2007ApJ...654..290S} associated to more energetic emissions.

\begin{figure}[htpb!]
\includegraphics[width=1.0\textwidth]{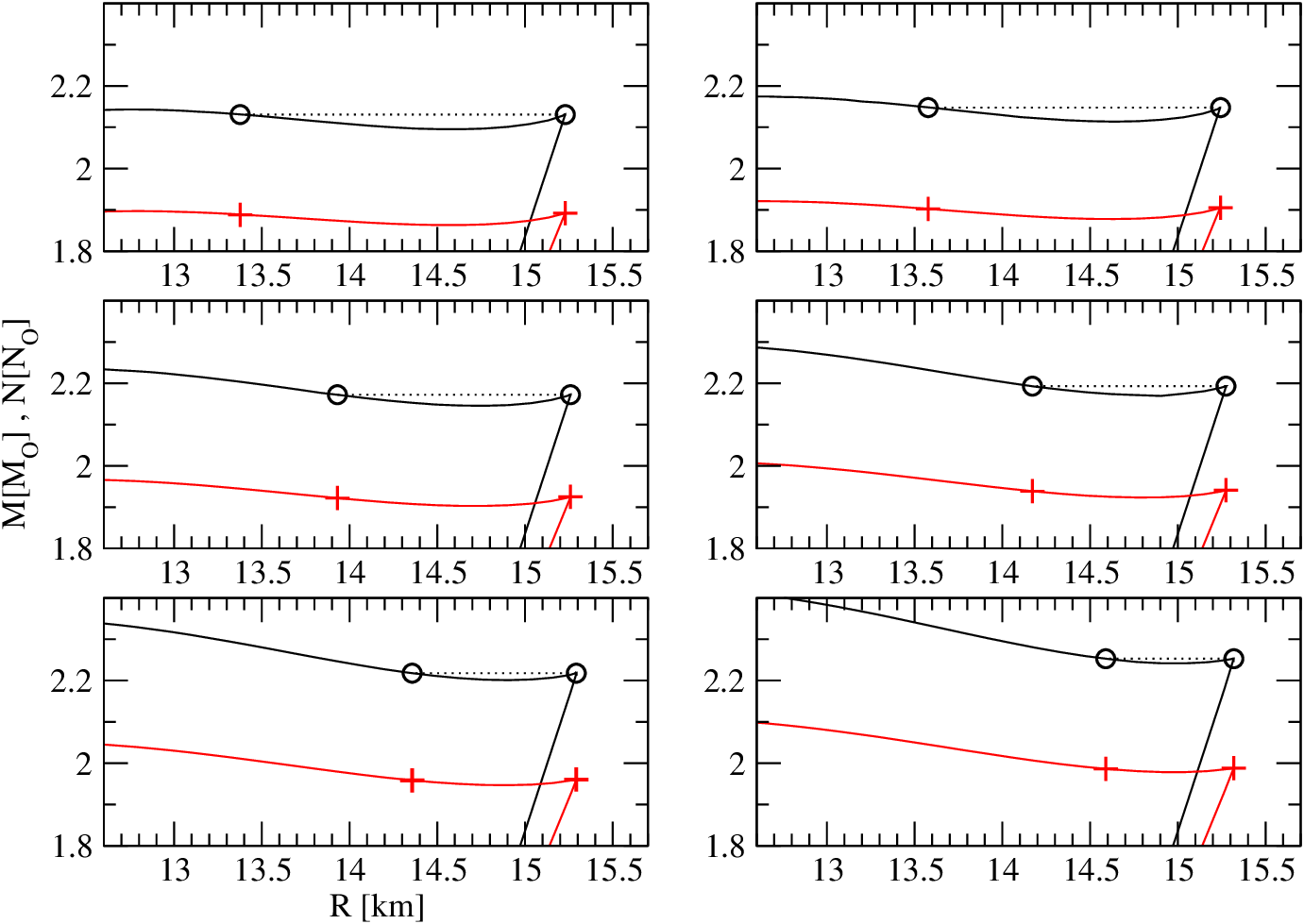}
\caption{Compact star configurations following an instability due to a strong first order phase transition. Baryonic mass and gravitational mass are given by the black and red curves, respectively. 
The models are characterized by $\eta_4$ values: 
0.0, 1.0, 3.0, 5.0, 7.0, 10.0 starting from the upper left corner down to the bottom right. 
Notice that the higher the $\eta_4$ value the higher the mass at the instability.}
\label{fig:MvsR}
\end{figure}

Neutron stars can suffer a dynamical collapse caused by a deconfinement phase transition in their cores (possibly leading to a {\it corequake}) via spin-down due to 
the continuous electromagnetic emission or by mass accretion.
Some early works estimated the energy reservoir for the typical ($\sim 1.4$M$_\odot$) NS mass \cite{1999JPhG...25..971D,2002NuPhS.113..268B,2004A&A...416..991A,Zdunik:2005kh},
however here we present a recent EoS model derived in \cite{2015A&A...577A..40B} that allows for the formation of a ``third family'' of compact stars near the maximum mass. 
Figure~\ref{fig:MvsR} shows both gravitational and baryonic masses for compact star configurations with a transition at high masses. We denote the two stars before and after
the transition as ``high mass twins'' because we assume that they bear the same baryonic mass while differing in their mass-energy quantity (binding energy). 
Although the detailed mechanism is at the moment being developed~\cite{Bejger:2015}, we can conjecture the following:
a deconfinement phase transition occurs via a corequake scenario in which a {\it high-mass} hadronic NS collapses into a hybrid compact
star disconnected from the former by a gap in the stable configurations (for a recent classification of hybrid stars, see \cite{Alford:2013aca}). 
The instability sets in by the pure hadronic NS after dipole-emission spin-down, or accretion-induced spin-up by matter from a companion.

For these energy bursts to occur it is important to note that in cold neutron stars
the temperatures in their interiors are well below the neutrino opacity temperature ($T_{\nu}\sim$1MeV) 
such that the free streaming scenario applies for neutrino propagation as opposed to the neutrino diffusion
mostly suitable for a gamma ray burst scenario involving a hot (proto) neutron star ($T_{\nu} <$ 1MeV).

A new aspect of this mechanism, going beyond the scenario of Falcke \& Rezzolla 
(a direct collapse of a magnetized NS to a black hole~\cite{2014A&A...562A.137F}) is the existence of a metastable state between
the initial and final states of a supramassive rotating neutron star SURON that could explain a double peak structure of the FRB's for which recently 
a case has been made~\cite{Champion:2015}. This metastate in the SURON process corresponds to an object on the third family branch of high mass hybrid stars, 
as found recently for microscopically motivated EoS~\cite{2014A&A...562A.137F,2015arXiv151105873A}.

\section{High-mass twin equation of state}

\begin{figure}[htpb!]
\centering
\includegraphics[width=1.0\textwidth]{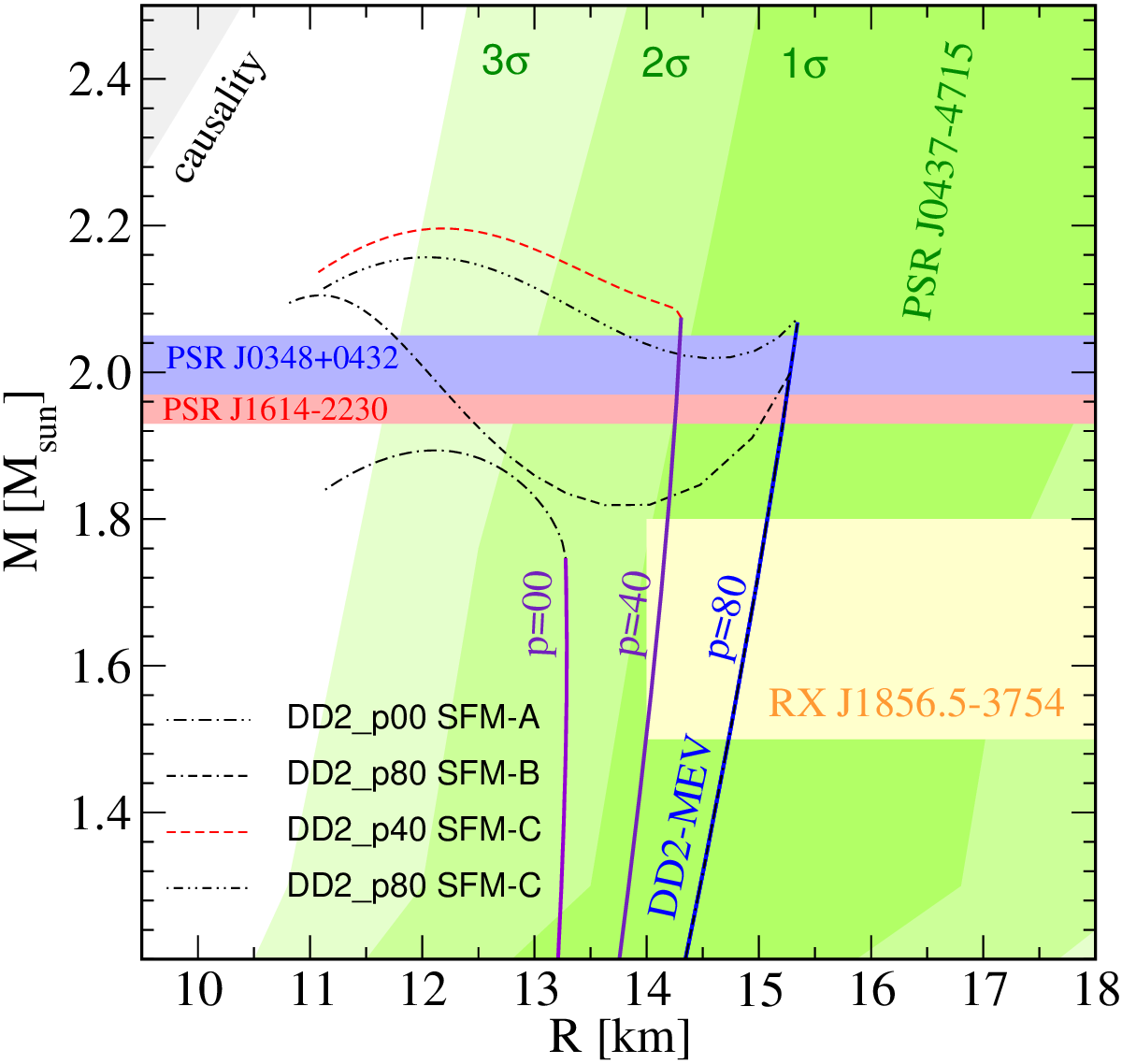}
\caption{String-Flip approach to the high-mass twins~\cite{2015arXiv151105873A} and constraints for mass and radius values. The green regions correspond to
1,2 and 3$\sigma$ values as reported in~\cite{2013ApJ...762...96B}.
The vertical bands around 2M$_{\odot}$ correspond to the most precisely massive NS measurements~\cite{2013Sci...340..448A,2010Natur.467.1081D}. Figure taken from~\cite{2015arXiv151105873A}.}
\label{fig:SF-Twins}
\end{figure}

We call neutron star \textit{twins} those stars with the same mass having different composition and thus different radius. They belong to different branches in the disconnected
mass-radius diagram. One of them is pure hadronic whereas the twin star is a hybrid containing quark matter in its 
core~\cite{1968PhRv..172.1325G,1981JPhA...14L.471K,2000NuPhA.677..463S,2000A&A...353L...9G,2015PhRvC..91e5808D}. The case of high mass twins is of great importance
because is a consequence of a critical point in the QCD phase diagram~\cite{2015arXiv150305576A, 2013arXiv1310.3803B}. Furthermore, they do not lead to many of the
modern issues of compact star physics as discussed in \cite{2015arXiv150303834B}: masquerades \cite{2005ApJ...629..969A}, the hyperon puzzle \cite{2003astro.ph.12446B}, 
and reconfinement~\cite{2011arXiv1112.6430L,2013A&A...551A..61Z}. 
Bayesian studies based on the most reliable observations provide a useful assessment 
of high mass twins identification~\cite{2014arXiv1412.8226A,2014JPhCS.496a2002B, Alvarez-Castillo:2015via,2015arXiv151105880A}.
The description of the EoS used here is as follows:

\begin{itemize}
 \item For hadronic matter we utilize the DD2 EoS with excluded volume correction (resulting from the internal compositions of nucleons
 produced by Pauli blocking effects of quarks~\cite{Ropke:1986qs}) that acts at suprasaturation densities and provides stiffening to this EoS.
 For neutron stars this means that the highest mass can be reached at a rather lower interior density values 
 and are characterized by large radii ($\simeq 13-15$ km, \cite{2013arXiv1310.3803B}).
 \item  Quark matter is described by a NJL EoS with multiquark interactions hNJL~\cite{2014EPJA...50..111B} featuring a coupling strength parameter
 in the vector channel of the 8-quark interaction $\eta_4$ bringing this quark EoS to a sufficient stiffening at high densities in order to support 
 the maximum observed star mass of 2~M$_{\odot}$. Alternatively, an equation of state based on the string-flip model~\cite{2015arXiv151105873A}
 captures the same features (excluded volume), but has the advantage of resulting in a broader range of radius difference between the twins 
 (in better agreement with the contrasting either large or short radius observations). The mass-radius diagrams for the string-flip compact stars is shown
 in figure~\ref{fig:SF-Twins}.
\end{itemize}
\section{Results and discussion}
The EoS of \cite{2015A&A...577A..40B} features a first order phase transitions 
leading to the instability of the NS right after the appearance of a small, dense quark core. 
When the NS reaches the maximum hadronic mass radial oscillations take over creating an instability~(see e.g. \cite{1986bhwd.book.....S}), which results into a
dynamical collapse. The criterion for unstable configurations is $\partial M/\partial n_c < 0$, where $n_c$ is the central density of the corresponding NS. 
In figure \ref{fig:MvsR} the configurations in between circles (maximum hadronic mass NS and its hybrid twin) that have a positive slope are thus forbidden. This is the origin
of the gap created by such prohibited configurations in the mass-radius diagram.
\begin{figure*}[htb!]
\begin{center}$
\begin{array}{cc}
\includegraphics[width=0.45\textwidth]{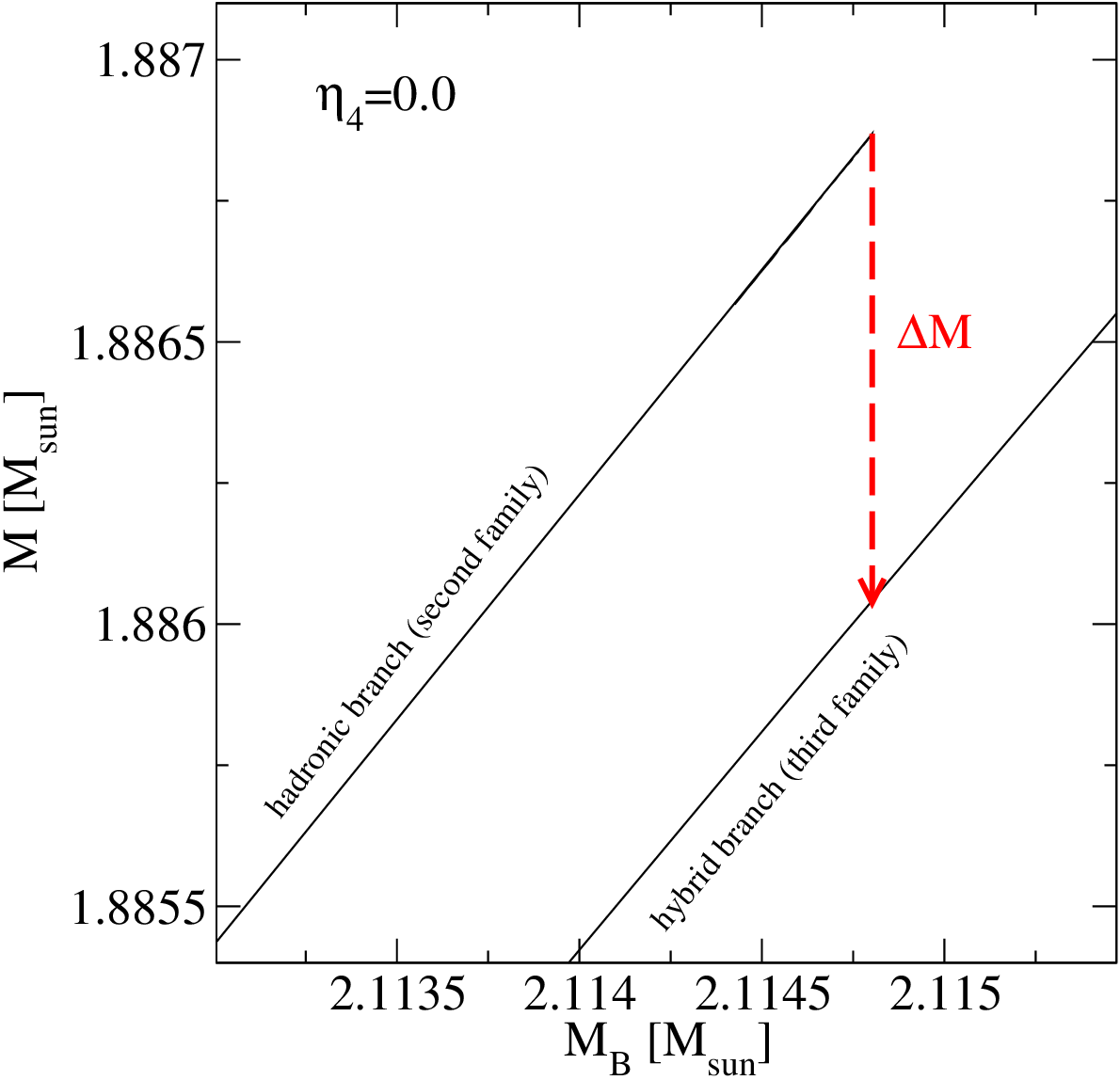} &
\includegraphics[width=0.45\textwidth]{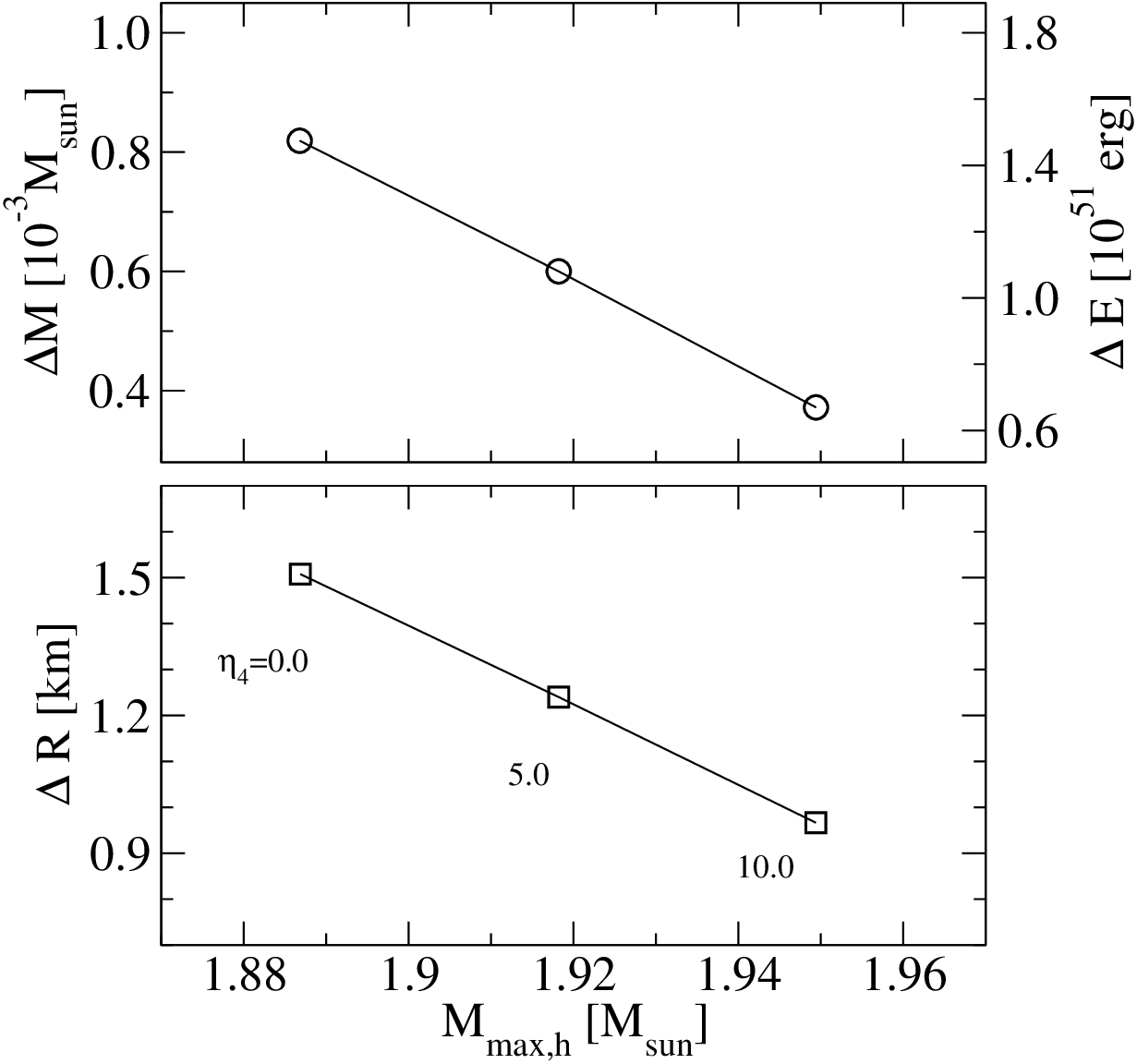}\\
\end{array}$
\end{center} 
\caption{\textit{Left.} Gravitational mass $M$ vs. baryonic mass $M_B$. $\eta_4=0.0$ is the value of the vector coupling in the chosen hNJL model shown here. The red line indicates the energy released in the transition between the maximum hadronic NS and the corresponding hybrid star mass twin. \textit{Right.} Mass difference $\Delta M$ (upper panel) and radius difference $\Delta R$ (lower panel) resulting 
from the dynamical NS collapse induced by a deconfinement phase transition for
a set of vector coupling parameters $\eta_4=0.0,5.0,10.0$ of the high-mass twin hNJL models. 
The corresponding energy release $\Delta E$ after the transition is indicated on the right side of the upper panel. Figures taken from~\cite{2015arXiv150608645A}.}
\label{Energy_Released}
\end{figure*}
Figure~\ref{Energy_Released} (left panel) shows the radius change $\Delta R$ in the transition, which 
is between $1$ and $1.5$ km 
for some coupling constant values.
The available energy $\Delta E$ following the transition 
equals the mass-energy difference $\Delta Mc^2$ between the 
initial and final configurations and is of about $10^{51}$ erg. See figure~\ref{Energy_Released} (right panel). 
The more realistic case of rotation configurations are being currently studied~\cite{Bejger:2015}. 

We conclude this work by drawing attention to the possibility of energetic emissions produced by a deconfinement phase transition in NS interiors most likely via FRB's, 
such as the case of the recent observations of FRB121002 \cite{Champion:2015} featuring a double peak light curve structure. We identify this double peak signal with the
metastable state in the dynamical collapse scenario of a SURON. Furthermore, it is of great importance to mention
that the NS EoS presented here can serve as an input to understand scenarios of cosmic ray generation via supernova explosions or NS mergers. 

\section{Acknowledgments}
D.E. Alvarez-Castillo would like to thank the organizers of \textit{The International Workshop on 
Quark Phase Transition in Compact Objects and Multimessenger Astronomy: 
Neutrino Signals, Supernovae and Gamma-Ray Bursts} for their support and hospitality. In particular, he acknowledges interesting and fruitful discussions with Alessandro Drago, 
Grzegorz Wiktorowicz, Christian Spiering, A.Kh.Khokonov and Igor Iosilevskiy during his stay at SAO and BNO.
The author also thanks David Blaschke, Stefan Typel, Mark Kaltenborn and Sanjin Beni\'c 
for providing the necessary EoS data sets for this work as well as Pawel Haensel, Michal Bejger and Leszek Zdunik for collaboration on the study of the 
high energy emission mechanism and dynamical collapse. 
D.E.A.-C. acknowledges support by the 
Heisenberg-Landau programme,
the Bogoliubov-Infeld programme and by the 
COST Action MP1304 ``NewCompStar''.

\newcommand{\aap}{A\&A}
\newcommand{\apj}{ApJ}
\newcommand{\prc}{Phys.~Rev.~C}
\newcommand{\physrep}{Phys.~Rep.~}
\newcommand{\apjl}{ApJ}
\newcommand{\nat}{Nature}
\newcommand{\apjs}{ApJS}
\newcommand{\prd}{Phys.~Rev.~D}
\newcommand{\ssr}{Space Sci. Rev.}
\newcommand{\apss}{Astrophysics and Space Science}
\newcommand{\procspie}{Proc. SPIE}

\clearpage

\end{document}